\documentclass[3p,times,procedia]{elsarticle}
\usepackage{nupha_ecrc}
\volume{00}

\firstpage{1}

\journalname{Nuclear Physics A}

\runauth{}

\jid{nupha}

\jnltitlelogo{Nuclear Physics A}

\usepackage{amssymb}
\usepackage{lineno}
\usepackage[figuresright]{rotating}

\begin{document}
\begin{frontmatter}



%% Instructions from Editor: Please use the following \dochead only in the preprint version (e-print arXiv etc.); 
%% use empty \dochead{} when submitting to Nuclear Physics A!
\dochead{XXVIIth International Conference on Ultrarelativistic Nucleus-Nucleus Collisions\\ (Quark Matter 2018)}
%% e.g. \dochead{17th International Conference on Dynamical Processes in Excited States of Solids}

\title{Energy and system dependence of nuclear modification factors of inclusive charged particles and identified light hadrons measured in p--Pb, Xe--Xe and Pb--Pb collisions with ALICE}


\author{Daiki Sekihata for the ALICE Collaboration}

\address{Hiroshima University, 1-3-1, Kagami-yama, Higashi-Hiroshima, Hiroshima, Japan}

\begin{abstract}
We report recent ALICE results on primary charged particle and neutral meson production in pp, p--Pb, Pb--Pb  and Xe--Xe collisions at LHC energies.
In this article, measurements of the nuclear modification factors $R_{\rm AA}$ of primary charged particles and of light neutral mesons in Pb--Pb, in Xe--Xe and in p--Pb collisions in a wide $p_{\rm T}$ range and different centrality classes are discussed.
We compare the nuclear modification factors obtained for different collision systems as a function of transverse momentum, collision centrality as well as charged particle multiplicity (${\rm d}N_{\rm ch} / {\rm d}\eta$). We also present comparison of experimental results to model calculations.
\end{abstract}


\end{frontmatter}

%\linenumbers

\section{Introduction}
\label{sec_intro}
Partons originating from initial hard scatterings lose their energy in the hot and dense QCD matter produced in ultra-relativistic heavy-ion collisions, which result in suppression of high $p_{\rm T}$ hadrons reported by several experiments \cite{Adcox:2001jp,Adler:2002xw,Aamodt:2010jd,Aad:2015wga,CMS:2012aa}, known as "jet quenching".
Light flavor hadrons are powerful probes to measure the suppression, because they can be measured in a wide transverse momentum ($p_{\rm T}$) range with high precision.
The modification of particle production at high $p_{\rm T}$ is quantified by the nuclear modification factor $R_{\rm AA}$, which is the ratio of particle yields in AA collisions to that in pp collisions at the same center-of-mass energy scaled by the number of binary nucleon-nucleon collisions $\langle N_{\rm coll} \rangle$,
\begin{equation}
R_{\rm AA} = \frac{{\rm d}N^{\rm AA}/{\rm d}p_{\rm T}}{ \langle N_{\rm coll} \rangle \times {\rm d}N^{\rm pp}/{\rm d}p_{\rm T}}
                   = \frac{{\rm d}N^{\rm AA}/{\rm d}p_{\rm T}}{ \langle T_{\rm AA} \rangle \times {\rm d}\sigma^{\rm pp}/{\rm d}p_{\rm T}}
\label{def_RAA}
\end{equation}
where $N^{\rm AA}$ and $N^{\rm pp}$ are particle yields in AA and pp collisions and $\sigma^{pp}$ is production cross section in pp collisions respectively.
The nuclear overlap function $\langle T_{\rm AA} \rangle = \langle N_{\rm coll} \rangle / \sigma_{\rm INEL}^{\rm pp}$ is determined from the Glauber model of the nuclear collision geometry and $\sigma_{\rm INEL}^{\rm pp}$ is the total inelastic nucleon-nucleon cross section \cite{Loizides:2017ack}.
%%%%%%%%%%%%%%%%%%%%%%%%%%%%%%%%%%%%%%%%%%%
\section{Nuclear modification factors in p--Pb and Pb--Pb collisions}
\label{sec_PbPb}
\begin{figure}[htbp]
\begin{minipage}{0.5\hsize}
  \centering
  \includegraphics[scale=0.38]{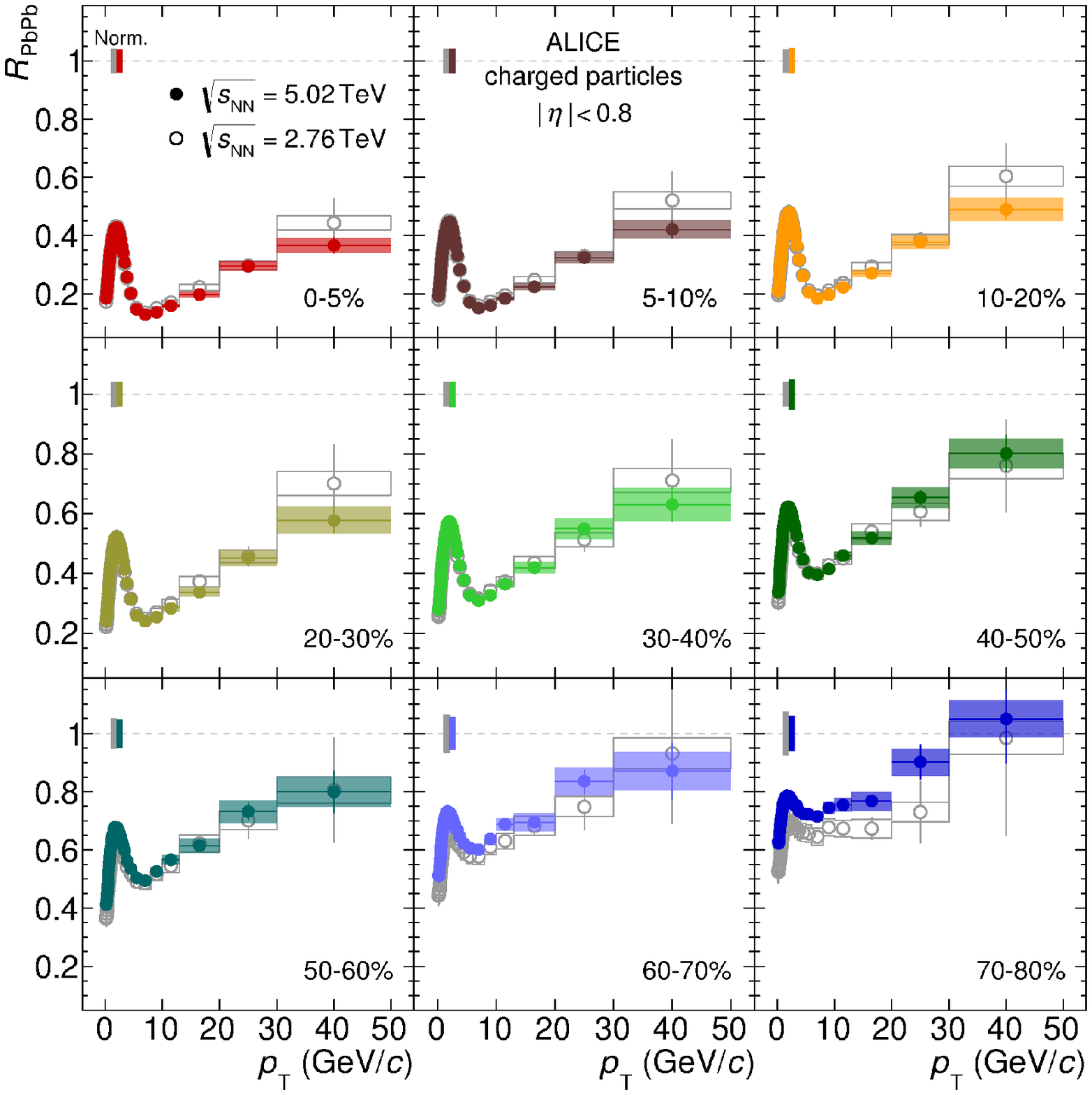}
 \end{minipage}
 \begin{minipage}{0.5\hsize}
  \centering
  \includegraphics[scale=0.40]{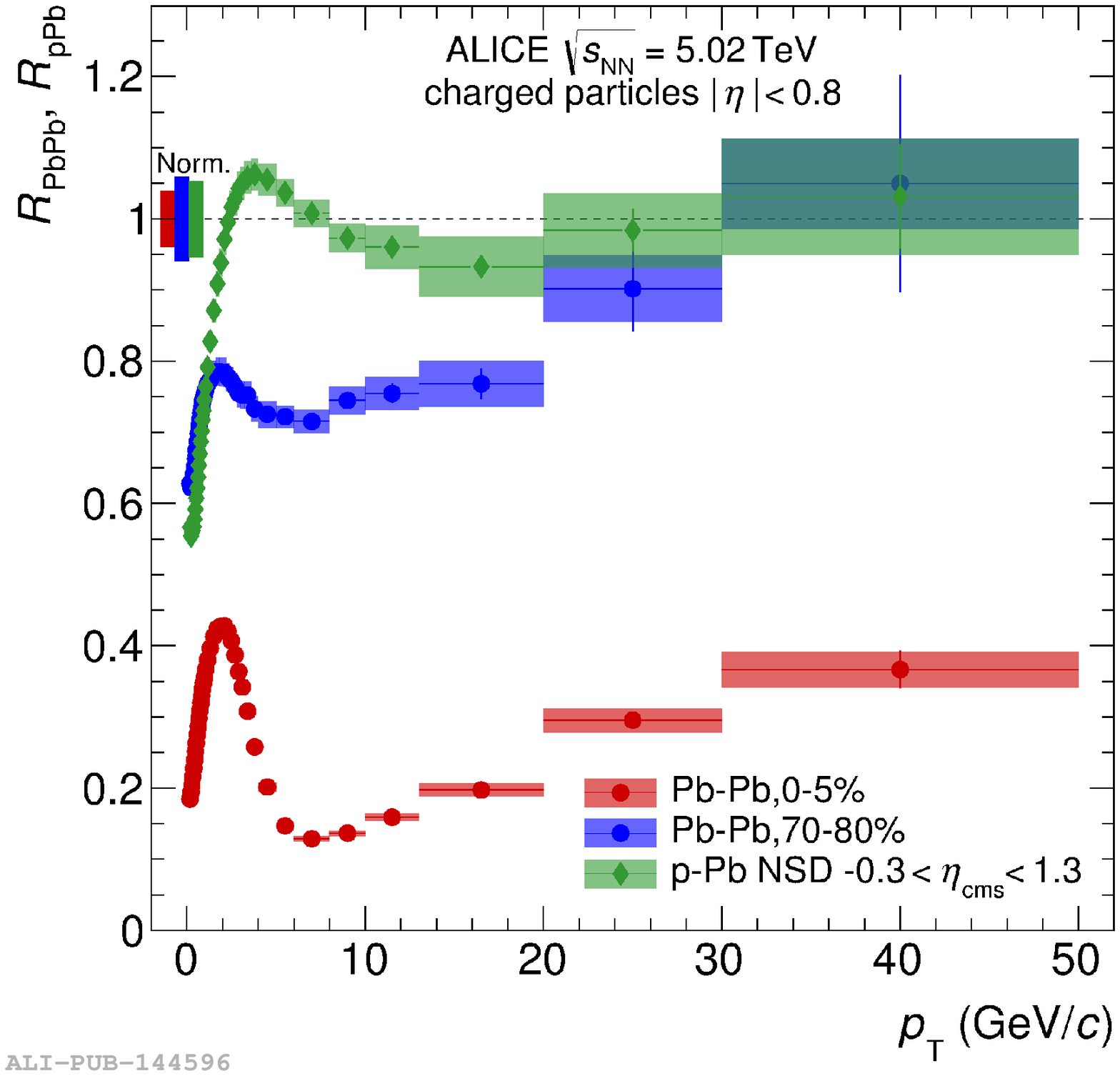}
  \end{minipage}
  \caption{Left : The transverse momentum dependence of the nuclear modification factor measured in Pb--Pb collisions for nine centrality classes. The new data at $\sqrt{s_{\rm NN}} = 5.02$ TeV are compared to the re-analyzed data at $\sqrt{s_{\rm NN}} = 2.76$ TeV \cite{Acharya:2018qsh}. Right : Nuclear modification factors of primary charged particles measured by ALICE in central (0-5\%) and peripheral (70-80\%) Pb--Pb collisions and in p--Pb collisions at $\sqrt{s_{\rm NN}} = 5.02$ TeV \cite{Acharya:2018qsh}.}
  \label{RAA_ch}
\end{figure}
\begin{figure}[htbp]
\begin{minipage}{0.5\hsize}
  \centering
  \includegraphics[scale=0.4]{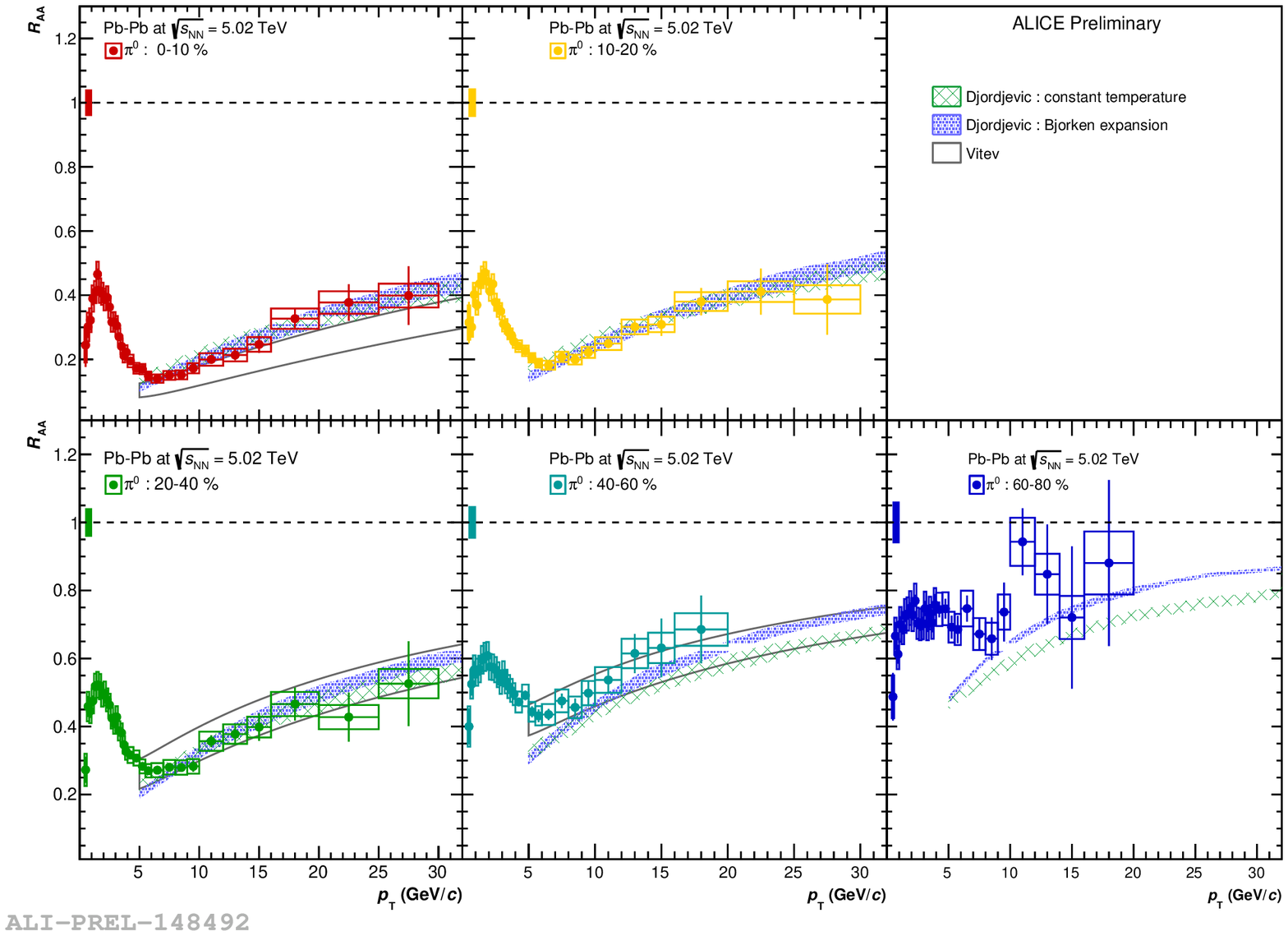}
 \end{minipage}
 \begin{minipage}{0.5\hsize}
  \centering
  \includegraphics[scale=0.32]{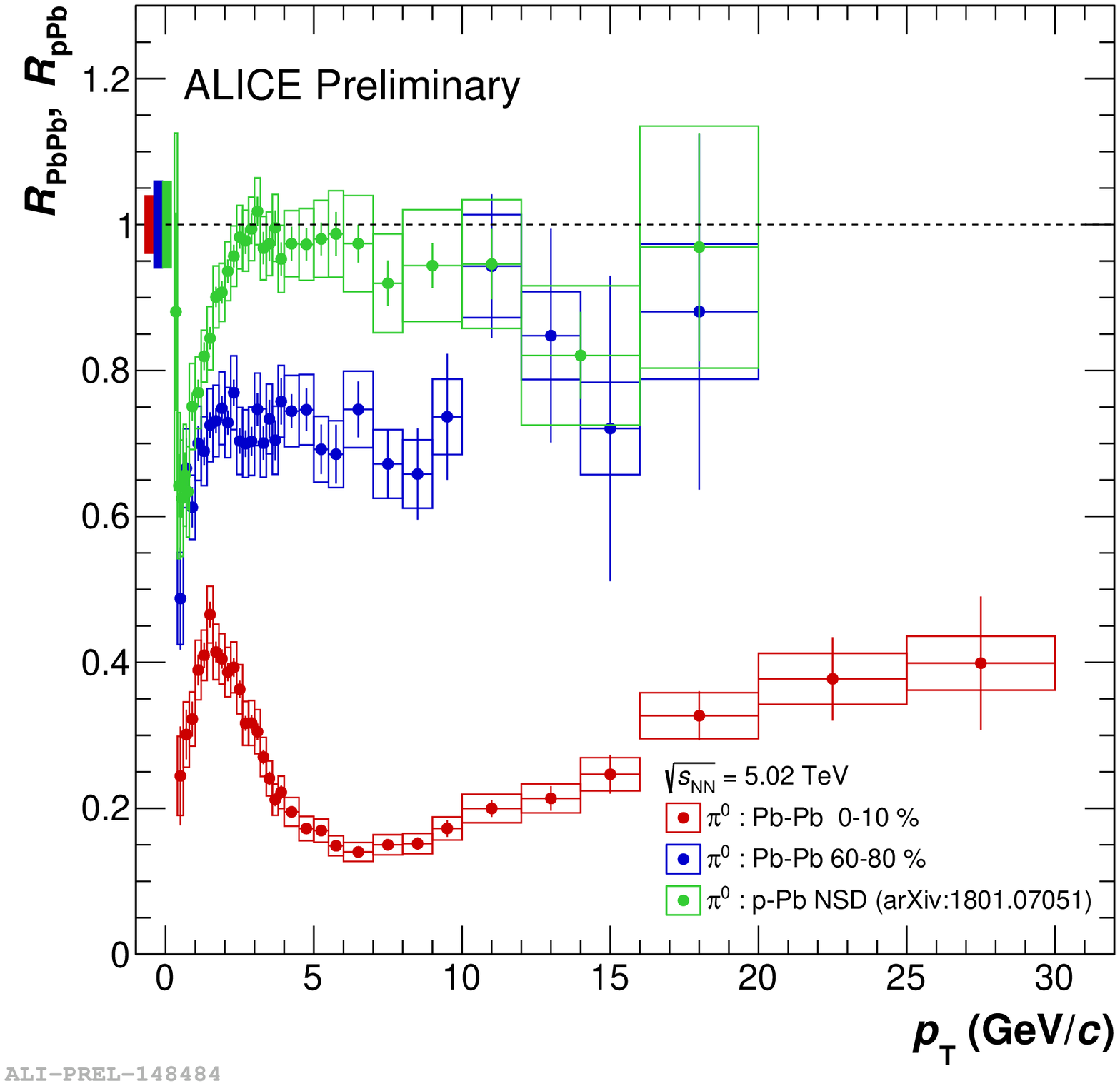}
  \end{minipage}
  \caption{Left : Nuclear modification factors of $\pi^{0}$ as a function of $p_{\rm T}$ in different centrality classes compared to theoretical models \cite{Zigic:2018smz,Zigic:2018ovr,Kang:2014xsa,Chien:2015vja} in Pb--Pb collisions at $\sqrt{s_{\rm NN}}$ = 5.02 TeV. Right : Nuclear modification factors of $\pi^{0}$ in central (0-10\%) and peripheral (60-80\%) Pb--Pb collisions and in minimum-bias p--Pb collisions at $\sqrt{s_{\rm NN}}$ = 5.02 TeV \cite{Acharya:2018hzf}.}
  \label{RAA_pi0}
\end{figure}

Charged particles are measured by the Inner Tracking system (ITS) and the Time Projection Chamber (TPC) in the central barrel of the ALICE apparatus \cite{Aamodt:2008zz,Abelev:2014ffa}.
Concerning identified particles, this article covers the production of neutral mesons while results for other identified particles are presented in \cite{FB_QM18}.
Neutral mesons are measured by two types of electro-magnetic calorimeters, which are EMCal and Photon Spectrometer (PHOS) as well as via photon conversions ($\gamma \rightarrow \rm{ee}$) in detector materials at the mid-rapidity.
The fully corrected $p_{\rm T}$ spectra of primary charged particles in the kinematic range of $0.15 < p_{\rm T} < 50$ GeV/$c$ and $|\eta| < 0.8$ have been measured in pp and Pb--Pb collisions at $\sqrt{s_{\rm NN}}$ = 2.76 and 5.02 TeV, and in p--Pb collisions at $\sqrt{s_{\rm NN}} = 5.02$ TeV with ALICE \cite{Acharya:2018qsh}.
A significant improvement of systematic uncertainties related to the tracking efficiency motivated the reanalysis of data in pp and Pb--Pb collisions at $\sqrt{s_{\rm NN}}$ = 2.76 TeV and p--Pb collisions at $\sqrt{s_{\rm NN}} = $ 5.02 TeV.
Fig.~\ref{RAA_ch} (left) shows nuclear modification factors of primary charged particles as a function of $p_{\rm T}$ in nine centrality classes.
Nuclear modification factors have a strong centrality dependence and are similar at the two collision energies.
As $p_{\rm T} $ spectra become harder at higher collision energy, this similarity in $R_{\rm AA}$ indicates larger parton energy loss in a hotter and denser QCD medium produced at the higher collision energy.
Fig.~\ref{RAA_ch} (right) shows $R_{\rm pA}$ compared to $R_{\rm AA}$ in 0-5\% and 70-80\% centrality classes for Pb--Pb collisions at $\sqrt{s_{\rm NN}}$ = 5.02 TeV.
$R_{\rm pA}$ is consistent with unity at high $p_{\rm T}$, which demonstrates that the strong suppression observed in central Pb--Pb collisions is related to the formation of a hot and dense QCD medium.
In addition, $R_{\rm AA}$ of $\pi^{0}$ have been measured in $0.4 < p_{T} < 30$ GeV/$c$ in Pb--Pb collisions at $\sqrt{s_{\rm NN}}$ = 5.02 TeV, as shown by Fig.~\ref{RAA_pi0} (left).
One of its advantages compared to inclusive charged particles is well defined fragmentation function for an identified hadron.
Thus, understanding the production of identified hadrons in pp collisions is also important baseline in this study.
The maximum $p_{\rm T}$ reach is extended to 30 GeV/$c$, compared to previous results at $\sqrt{s_{\rm NN}}$ = 2.76 TeV \cite{Abelev:2014ypa,Acharya:2018yhg}.
$R_{\rm AA}$ of $\pi^{0}$ at two collision energies is also similar, as well as that of primary charged particles.
Furthermore, the behavior of $R_{\rm pPb}$ at high $p_{\rm T}$ is the same as that of primary charged particles (Fig.~\ref{RAA_pi0} (right)).
We also present a comparison of the measured $R_{\rm AA}$ of $\pi^{0}$ with theoretical models in Fig.~\ref{RAA_pi0} (left).
The model calculations by Djordjevic et al. \cite{Zigic:2018smz,Zigic:2018ovr} and Vitev et al. \cite{Kang:2014xsa,Chien:2015vja} give a quantitatively good description of $p_{\rm T}$ and centrality dependence of $R_{\rm AA}$.
%%%%%%%%%%%%%%%%%%%%%%%%%%%%%%%%%%%%%%%%%%%%%%%%%%%%%%
\section{Nuclear modification factors in Xe--Xe collisions}
\label{sec_XeXe}
\begin{figure}[htbp]
\begin{minipage}{0.5\hsize}
  \centering
  \includegraphics[scale=0.4]{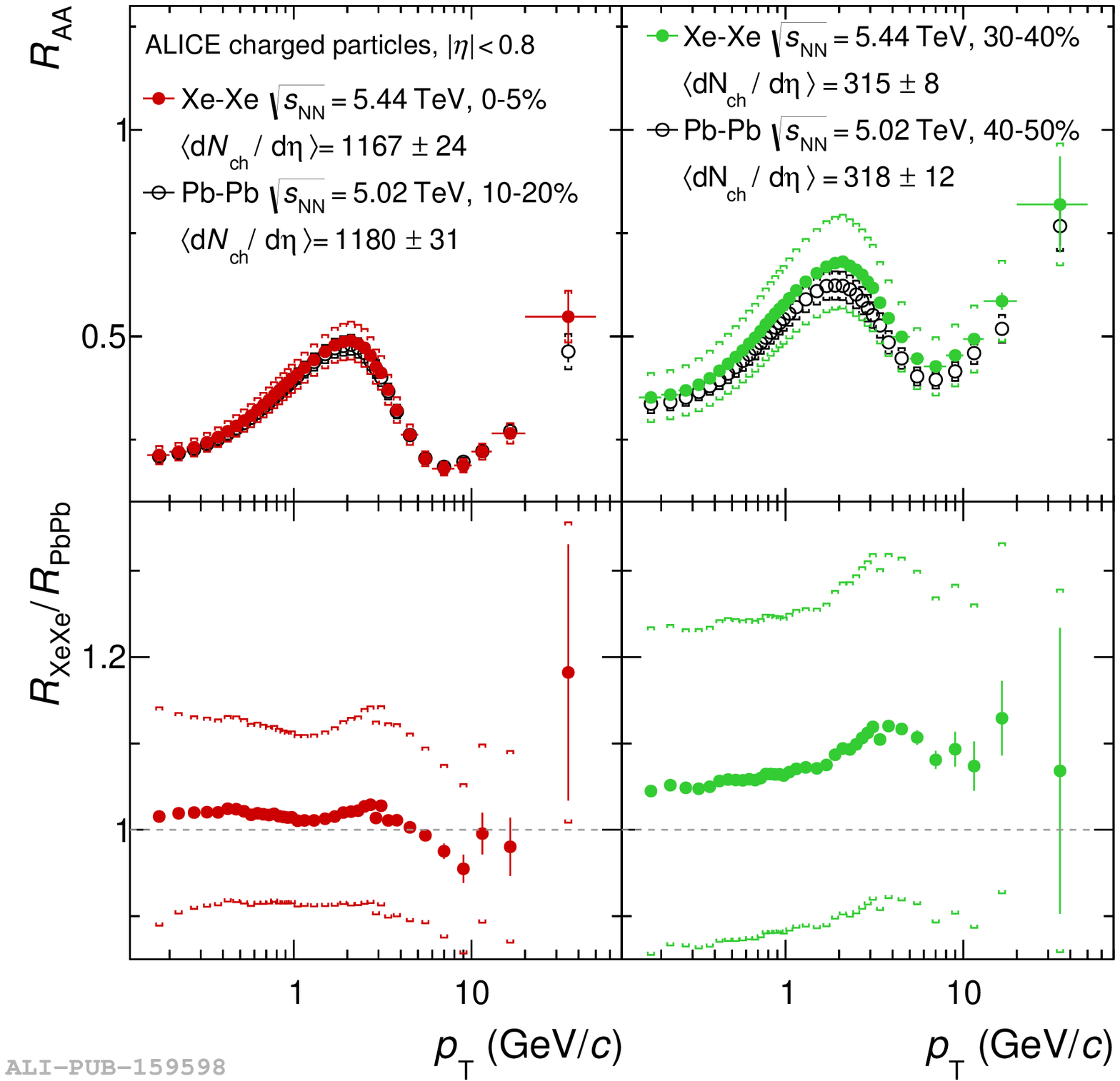}
 \end{minipage}
 \begin{minipage}{0.5\hsize}
  \centering
  \includegraphics[scale=0.3]{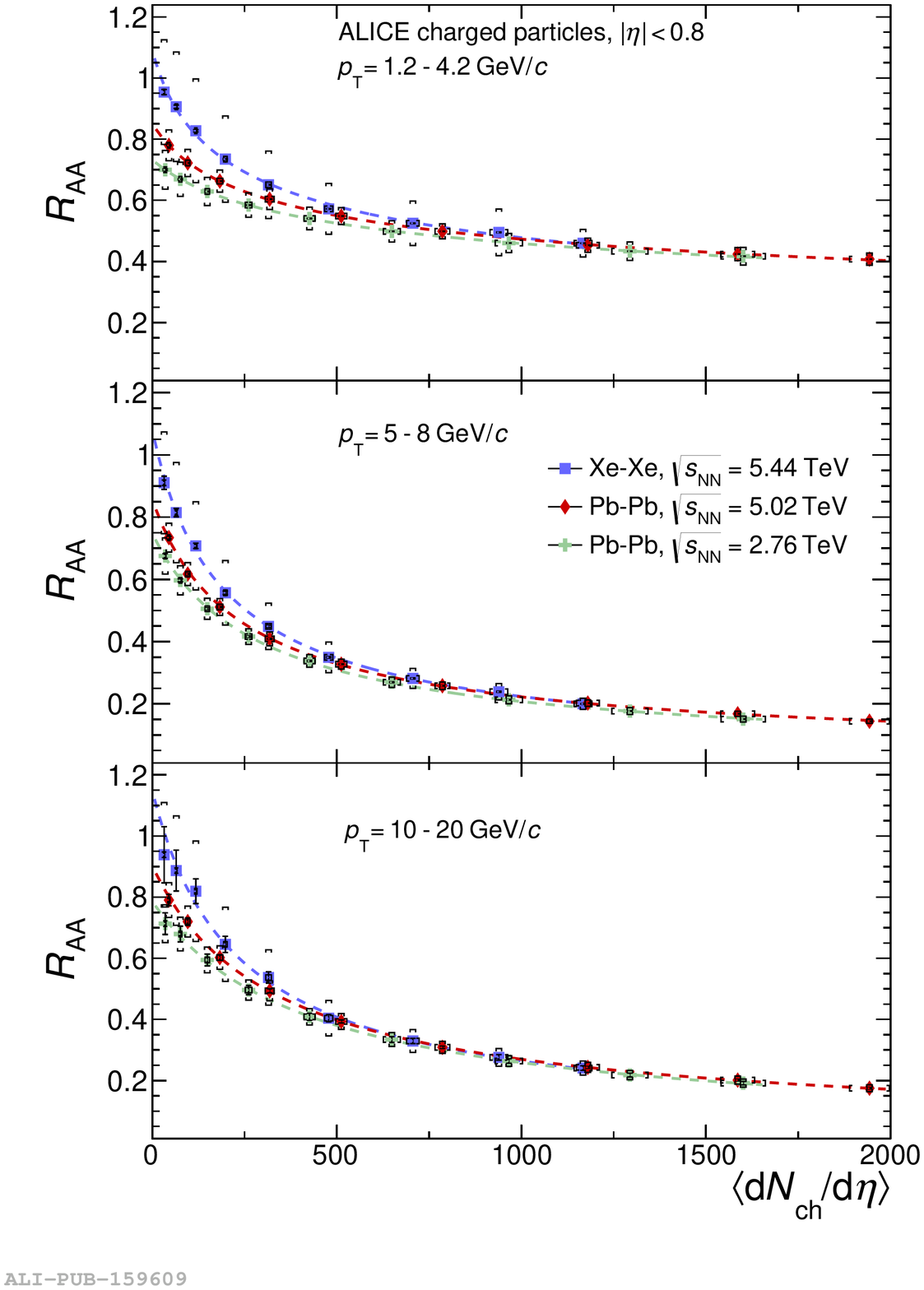}
  \end{minipage}
  \caption{Left : The nuclear modification factors in Xe--Xe collisions and Pb--Pb collisions for similar values in $\langle {\rm d}N_{\rm ch} / {\rm d}\eta \rangle$ for the 0-5\% and 30-40\% Xe--Xe centrality classes \cite{Acharya:2018eaq}. Right : Comparison of the nuclear modification factor in Xe--Xe and Pb--Pb collisions integrated over identical regions in $p_{\rm T}$ as a function of $\langle {\rm d}N_{\rm ch} / {\rm d}\eta \rangle$ \cite{Acharya:2018eaq}.}
  \label{RAA_ch_XeXe}
\end{figure}

The fully corrected $p_{\rm T}$ spectra of primary charged particles have been measured in Xe--Xe collisions at $\sqrt{s_{\rm NN}}$ = 5.44 TeV \cite{Acharya:2018eaq} in the same kinematic range as in Pb--Pb collisions.
The pp reference is interpolated from $p_{\rm T}$ spectra in pp collisions at $\sqrt{s}$ = 5.02 and 7 TeV by using a power-law parameterization in order to determine nuclear modification factors.
The nuclear modification factor exhibits a strong centrality dependence with a minimum at $p_{\rm T}$ = 6-7 GeV/$c$ and an almost linear rise in the higher $p_{\rm T}$ region.
In particular, the yield in the most central collisions (0-5\%) is suppressed by a factor of about 6 at minimum with respect to the scaled pp reference.
$R_{\rm AA}$ reaches a value of 0.6 at the highest $p_{\rm T}$ interval of 30-50 GeV/$c$.
A similar characteristic $p_{\rm T}$ dependence of $R_{\rm AA}$ is observed in both Xe--Xe and Pb--Pb collisions.
Fig.~\ref{RAA_ch_XeXe} (left) shows the comparison of $R_{\rm AA}$ between Xe--Xe and Pb--Pb collisions for the same charged particle multiplicity $\langle {\rm d}N_{\rm ch} / {\rm d}\eta \rangle$ ranges and their ratios.
In the most central Xe--Xe collisions (0-5\%), the nuclear modification is consistent with that in 10-20\% central Pb--Pb collisions over the entire $p_{\rm T}$ range.
While a similar $R_{\rm AA}$ is found for comparable charged particles multiplicity $\langle {\rm d}N_{\rm ch} / {\rm d}\eta \rangle$, the respective mean number of participants $\langle N_{\rm part}\rangle$ are significantly different.
$\langle N_{\rm part}\rangle$ is $236 \pm 2$ in the 0-5\% centrality class in Xe--Xe, but $263 \pm 4$ in the 10-20\% centrality class in Pb--Pb collisions \cite{GlauberXeXe,Acharya:2018hhy}.
In the 30-40\% Xe--Xe (40-50\% Pb--Pb) centrality class, agreement of $R_{\rm AA}$ is also found within uncertainties at similar $\langle N_{\rm part}\rangle$ of $82 \pm4$ ($86 \pm 2$).
A detailed comparison of nuclear modification factors as a function of $\langle {\rm d}N_{\rm ch} / {\rm d}\eta \rangle$ in Xe--Xe and Pb--Pb collisions for three selected $p_{\rm T}$ regions is shown in Fig.~\ref{RAA_ch_XeXe} (right).
A remarkable similarity in $R_{\rm AA}$ between Xe--Xe and Pb-Pb collisions is observed for $\langle {\rm d}N_{\rm ch} / {\rm d}\eta \rangle > 400$.
In a simplified radiative energy loss model, the average energy loss is proportional to the energy density and to the square of path length $L$ in the medium $\langle \Delta E \rangle \propto \varepsilon \cdot L^{2}$ \cite{dEnterria:2009xfs}.
Therefore, the comparison of the measured $R_{\rm AA}$ in two different collision systems can provide insight into path length dependence of medium-induced parton energy loss.
%%%%%%%%%%%%%%%%%%%%%%%%%%%%%%%%%%%%%%%%%%%%%%%%%%%%%%%
\section{Summary}
\label{sec_summary}
In summary, nuclear modification factors of primary charged particles and $\pi^{0}$ mesons have been measured in a wide $p_{\rm T}$ range at mid-rapidity in various centrality classes and different collision systems and energies.
$R_{\rm AA}$ shows similar value at two collision energies, indicating the presence of a hotter and dense QCD matter at higher collision energy.
$R_{\rm pA}$ which is consistent with unity at high $p_{\rm T}$ demonstrates that the strong suppression observed in central Pb--Pb collisions is related to formation of hot and dense QCD medium.
The measured $R_{\rm AA}$ is compared to theoretical models.
The model calculations by Djordjevic et al. \cite{Zigic:2018smz,Zigic:2018ovr} and Vitev et al. \cite{Kang:2014xsa,Chien:2015vja} give a quantitatively good description of the data.
A similar $R_{\rm AA}$ is observed in Xe--Xe collisions at $\sqrt{s_{\rm NN}}$ = 5.44 TeV and Pb--Pb collisions at $\sqrt{s_{\rm NN}}$ = 2.76 and 5.02 TeV in centrality classes corresponding to similar charged particle multiplicities.
This comparison of $R_{\rm AA}$ in two collision systems can provide insight into the path length dependence of medium-induced parton energy loss.


%% References

\bibliographystyle{elsarticle-num}
\bibliography{ref}


%\begin{thebibliography}{00}
%%
%%%% \bibitem must have the following form:
%%%%   \bibitem{key}...
%
%\bibitem{}
%
%\end{thebibliography}

\end{document}